\documentclass[12pt,notitlepage,amsmath,amssymb,floatfix,superscriptaddress]{revtex4}

\usepackage{amsmath}
\usepackage{amssymb}
\usepackage[dvips]{graphicx}
\usepackage{epsfig}
\usepackage{tabularx}
\usepackage{color}

\begin{document}

\title{Phonons mimicking doubly special relativity kinematics}

	\author{Francesco Marino}
	\affiliation{CNR-Istituto Nazionale di Ottica, Via Sansone 1, I-50019 Sesto Fiorentino (FI), Italy.}
	\affiliation{INFN, Sezione di Firenze, Via Sansone 1, I-50019 Sesto Fiorentino (FI), Italy}

\date{\today}

\begin{abstract}
Collective excitations (phonons) in barotropic, irrotational, inviscid fluids exhibit an effective Lorentz invariance, where the sound speed plays the role of the invariant speed of light in special relativity. By carefully selecting the interaction potentials, we explicitly construct two hydrodynamic models in which phonons obey doubly special relativistic kinematics, with the analogue Planck scale emerging from non-Newtonian behaviour at high energies. Specifically, we demonstrate that elastic storage leads to an approximate realization of Amelino-Camelia's scenario, while the Magueijo-Smolin model naturally emerges in the presence of elastic restoring forces.

\end{abstract}

\maketitle

\section{Introduction}

Different candidate theories of quantum gravity, such as loop quantum gravity \cite{rovelli}, string theory \cite{veneziano,gross}, as well as quantum gravity phenomenological approaches \cite{garay}, suggest the existence of a minimal length, usually identified with the Planck length $L_p$. If such length represents a fundamental kinematic property of spacetime, as is widely believed, it must remain the same value for all inertial observers, which appears in contrast with the FitzGerald-Lorentz contraction, a key consequence of special relativity. This fact has motivated the proposal of modified versions of special relativity, incorporating $L_p$ as an additional observer-independent scale (besides the light speed, $c$), or equivalently, a maximum energy scale $E_p=h c/L_p$.

Theories based on these assumption are known as Doubly Special Relativity (DSR) theories and involve a modification of the relativistic energy-momentum relation associated to deformed Lorentz transformations \cite{amelino1,amelino2}. Although in these theories the Lorentz symmetry is not preserved at the Planck scale, this doesn't imply the loss of covariance and the emergence of a preferred class of inertial observers \cite{amelino2}. 
The deformation becomes negligible in the low-energy limit, so that a DSR model reduces to standard special relativity, exactly as special relativity reduces to Galilei relativity in the low-velocity limit. While various DSR theories have been developed, two prominent examples are the models by Amelino-Camelia \cite{amelino3} and Magueijo-Smolin \cite{ms}, commonly referred to as DSR1 and DSR2.

The DSR1 energy-momentum relation at $\mathcal{O}(E^3/E_p^3)$ is 
\begin{equation}
E^2 \simeq c^2 p^2 \left(1+ \frac{E}{E_p} \right) + m^2 c^4
\label{eq1}
\end{equation}
where $E$ is the total energy of a particle with momentum $p$ and rest mass $m$.
At this order of approximation, Eq. (\ref{eq1}) can be directly compared with the DSR2 dispersion relation
\begin{equation}
E^2 = c^2 p^2 + m^2 c^4\left(1 - \frac{E}{E_p} \right)^2
\label{eq2}
\end{equation}
Both models include an energy-dependent correction term which becomes significant when $E \sim E_p$. However, in the nonrelativistic limit, the deformation in Eq. (\ref{eq1}) can be understood as modifying the particle's inertial mass while keeping the rest mass unchanged \cite{jafari}, whereas in Eq. \ref{eq2}, the opposite occurs \cite{coreddu}. In this sense, the two theories seems to be complementary to each other. 

While DSR dispersion relations are motivated by the need to introduce an additional invariant scale inspired by quantum gravity considerations, a natural question is whether similar dispersion relations can emerge also in contexts beyond the realm of gravitational physics, such as the kinematics of quasiparticles in many-body systems. 

The so-called analogue gravity program \cite{agrev} has demonstrated the potential of condensed-matter systems to simulate important phenomena of quantum field theory in curved spacetime \cite{natrev}, culminated in the recent experimental evidence of Hawking radiation \cite{philbin,weinfurtner2011,euve2016,steinhauer1,steinhauer2,drori,kolobov,shi,tamura} and rotational superradiance \cite{torres,cromb,braidotti}. Of particular relevance to this work, it has been shown that modified dispersion relations with an emergent low-energy Lorentz symmetry are a common characteristic of collective excitations in both classical and quantum fluids \cite{natrev}. The dispersion relation of surface gravity waves in shallow water \cite{rousseaux} and the Bogoliubov dispersion relation in atomic \cite{bogoliubov,reviewBEC} and photonic Bose gases \cite{fontaine18,falque} are well-known examples.
In these systems, the surface wave velocity and the sound speed respectively act as analogues of the invariant speed of light in special relativity, and the scale at which the effective Lorentz symmetry is broken (the acoustic analogue of the Planck scale) is related to the underlying microscopic physics of the system. In ideal classical fluids such a scale corresponds to the mean intermolecular distance and in superfluids to the minimal length over which the quantum fluid maintains its quantum coherence (coherence length).

However, in all the hydrodynamic systems mentioned above (and, to our knowledge, in all analogue gravity models), modifications to the dispersion relation are encoded in higher-order momentum terms, resulting in an explicit breakdown of Lorentz symmetry near the Planck scale. In DSR models, the Planck energy is instead introduced as an additional invariant scale and the Lorentz transformations are deformed accordingly to leave invariant the deformed dispersion relation.

As shown by Unruh in his seminal work in 1981 \cite{unruh} and later further examined by Visser \cite{visser}, a barotropic, inviscid, and irrotational flow provides the minimal framework for the emergence Lorentz-invariant phonon dispersion relation. In this Letter, we demonstrate that by adding repulsive dipolar interactions and non-Newtonian (viscoelastic) effects, we can explicitly construct fluid models in which phonons obey doubly special relativistic kinematics. Specifically, we show that storage viscosity leads to an approximate realization of Amelino-Camelia's scenario, while the Magueijo-Smolin model naturally emerges in the presence of elastic restoring forces.

\section{Acoustic analogue models of DSR}

We start writing the modified Klein-Gordon equation corresponding to the dispersion relation (\ref{eq1}). Replacing energy and momentum with the usual quantum mechanical operators $E=i\hbar \partial_{T}$ and ${\bf p}=-i\hbar {\bf \nabla}$ and the speed of light $c$ with the sound speed $c_s$ we obtain
\begin{equation}
\partial_{TT}^2 \psi_1= c_s^2 \nabla^2 \psi_1 - \Omega_0^2 \psi_1 - i \gamma \partial_T \nabla^2 \psi_1 \, ,
\label{eq3}
\end{equation}
where $\Omega_0=m c_s^2/\hbar$ is the particle rest frequency, $\gamma=\hbar c_s^2/E_p$ and $E_p$ is a sonic analogue of the Planck energy.

To construct a fluid model in which phonons obey Eq. (\ref{eq3}), we begin by recalling that, in the absence of mass sources or sinks, the fluid density $\rho$ and the flow ${\bf v}$ are related by the continuity equation 
\begin{equation}
\partial_t \rho + {\bf \nabla}\cdot (\rho {\bf v}) = 0 \, ,
\label{eq4} 
\end{equation}
which expresses mass conservation. Phonons are defined by first-order fluctuations $\{\rho_1, {\bf v_1}\}$ of the quantities describing the background fluid flow $\{\rho_0, {\bf v_0}\}$, where the total density and velocity fields are given by $\rho = \rho_0 + \rho_1$ and ${\bf v} = {\bf v_0} + {\bf v_1}$, with $\rho_1 \ll \rho_0$ and ${\bf v_1} \ll {\bf v_0}$. Substituting these expressions into Eq. (\ref{eq2}) we obtain the linearized continuity equation 
\begin{equation}
\partial_t \rho_1 + {\bf \nabla}\cdot (\rho_0 {\bf v_1} + \rho_1 {\bf v_1}) = 0 
\label{eq5} 
\end{equation}
which provides the first relationship between density and velocity fluctuations.

Under the assumption of irrotational flow, the fluid velocity field ${\bf v}$ can be described in terms of a scalar potential $\psi$, such that ${\bf v}={\bf \nabla} \psi$. This allows us to identify the variable $\psi_1$ in Eq. (\ref{eq3}) with the fluctuations of the (yet-to-be-determined) velocity potential $\psi$, such that ${\bf v_1}= {\bf \nabla} \psi_1$. 

Since $\psi_1$ represents the fluctuations of a hydrodynamic velocity potential, it must satisfy a linearized Euler equation. Therefore, the next step is to properly define the forces acting on the fluid so that Eq. (\ref{eq3}) can be rewritten in the form of a linearized Euler equation
\begin{equation}
\partial_t \psi_1 + {\bf v_0} \cdot {\bf \nabla} \psi_1 =  - \frac{P_1}{\rho_0} - \phi_1 - \Phi_1 \, ,
\label{eq6} 
\end{equation}
where $P_1$ denotes fluctuations of the bulk pressure $P$, while $\phi_1$ and $\Phi_1$ represent two linearized interaction potentials, which will be specified later. The coupled equations (\ref{eq5}-\ref{eq6}) univocally determine phonon dynamics.

The bulk pressure $P$ arising from local (contact) interactions between fluid particles, is a necessary condition for the propagation of sound. The barotropic assumption $P=P(\rho)$ guarantees that the speed of sound defined as $c_s^2=\partial P (\rho_0)/\partial \rho$ is a local function of the background density only. A position- and time-independent sound speed throughout the fluid is a necessary condition for a global Lorentz invariant dispersion relation. In standard BECs with purely local interactions, this amounts to requiring that the product of density times interaction strength is constant through the condensate. Since here we focus on reproducing DSR dispersion relations in a flat acoustic spacetime (i.e., a spatially homogeneous background), we will assume, without any loss of generality, both a constant density $\rho_0$ and a constant interaction strength.

Defining the comoving derivative $\partial_T = \partial_t + {\bf v_0} \cdot {\bf \nabla}$, the linearised continuity and Euler equations become
\begin{eqnarray}
\partial_T \rho_1 + \rho_0 \nabla^2 \psi_1 = 0 \label{eq7a}\; \\
\partial_T \psi_1 = - \frac{P_1}{\rho_0} - \phi_1 - \Phi_1 \label{eq7b} %\; \\
\label{eq7}
\end{eqnarray}
We now derive and analyze the terms on the right-hand side of Eq. (\ref{eq7b}), which are essential for obtaining the desired dispersion relation.

Expanding the bulk pressure $P(\rho_0 + \rho_1) \approx P(\rho_0) + \frac{\partial P(\rho_0)}{\partial \rho} \rho_1$ the bulk pressure fluctuations can be written as
\begin{equation} 
P_1=c_s^2 \rho_1
\label{eq7c}
\end{equation}

It is well known that for $\phi_1=\Phi_1=0$, Eqs. (\ref{eq7}) yield the dispersion relation of relativistic massless particles $E = c_s p$ \cite{landau}, where $E = E^{'} - {\bf p} \cdot \bf{v_0}$ is calculated in the reference frame comoving with the flow and $E^{'}$ is the phonon energy in the laboratory frame. Phonons always travel at the speed of sound relative to the flow and therefore they behave as massless collective excitations. While this is the case for most fluids, some hydrodynamic systems such as coupled-condensates \cite{silke-visser} (see also the recent experiment \cite{ferrari22}) or fluids with suitably shaped interactions \cite{eg2,marino2019,falque} can support massive excitations and even allow for the coexistence of both massive and massless modes.
In the simplest scenario, the inclusion of repulsive nonlocal (dipolar) interactions is sufficient to impart a finite rest mass to phonons\cite{marino2019,ciszak2021,marino2024}. Detailed models of excitation dynamics in dipolar Bose-Einstein condensates, with applications to analogue gravity, have recently been reported in Refs. \cite{holanda1,holanda2}. Here, we assume that fluctuations in the dipolar potential are driven by fluid density fluctuations through the Poisson equation
\begin{equation}
\nabla^{2} \phi_1 = - \beta \rho_1
\label{eq8}
\end{equation}
where $\beta >0$ accounts for the strength of the interactions. %For constant background density $\rho_0$, Eq. (\ref{eq4}) implies ${\bf \nabla}\cdot {\bf v_0} =0$. 
For a spatially-homogeneous flow ${\bf v_0}$, $\partial_T \nabla^2 = \nabla^2 \partial_T$ and hence by applying $\partial_T$ to Eq. (\ref{eq8}) and using (\ref{eq7a}) we obtain 
\begin{equation}
\partial_T \phi_1 = \beta \rho_0 \psi_1
\label{eq9}
\end{equation}
Since $\beta$ is assumed to be constant, a constant density is necessary to ensure a position-independent rest frequency for excitations within the fluid.

We have now all the ingredients to recast Eq. (\ref{eq3}) in the form of the linearized Euler equation (\ref{eq7b}). We first notice that the linearized continuity equation (\ref{eq7a}) dictates
\begin{equation}
c_s^2 \nabla^2 \psi_1 = - \frac{c_s^2}{\rho_0}\partial_T \rho_1 = -\frac{1}{\rho_0}\partial_T P_1 \, .
\label{eq10} 
\end{equation}
Substituting Eqs. (\ref{eq8})-(\ref{eq10}) in (\ref{eq3}) and defining the rest frequency $\Omega_0=\sqrt{\beta \rho_0}$ and $\Phi_1=i \gamma \nabla^2 \psi_1$ we obtain
\begin{equation}
\partial_{TT}^2 \psi_1 = -\frac{1}{\rho_0}\partial_T P_1  - \partial_T \phi_1 - \partial_T \Phi_1 \, ,
\label{eq11}
\end{equation}
which reduces to (\ref{eq7b}) after integration with respect to $T$. 

At the fully nonlinear level the corresponding Euler equation for momentum conservation and the Poisson equation describing the dipolar interactions read
\begin{eqnarray}
\partial_t {\bf v} +({\bf v} \cdot {\bf \nabla}) {\bf v}= - \frac{\nabla P}{\rho} - \nabla \phi - i \gamma \nabla^2 {\bf v} \label{eq12a}\; \\
\nabla^{2} \phi = -\beta \rho \label{eq12b}
\end{eqnarray}
For $\gamma=0$, Eqs. (\ref{eq12a}-\ref{eq12b}), together with the continuity equation (\ref{eq7b}) reduce to the Navier-Stokes equations for an irrotational fluid with both contact and dipolar interactions. 

For finite $\gamma$, the term $i \gamma \nabla^2 {\bf v}$ corresponds to a reactive, non-dissipative effect (momentum storage), in contrast to conventional viscosity $\eta \nabla^2 {\bf v}$ associated with momentum dissipation \cite{landau}. In rheology, this imaginary viscosity is referred to as the storage modulus (or storage viscosity) and quantifies the energy stored during deformation in a viscoelastic fluid \cite{chhabra}. These fluids can temporarily store deformation energy, which can later be released to restore the fluid micro-structure.

Since any realistic viscoelastic fluid exhibits both dissipative and elastic behavior, the viscosity coefficient appering before $\nabla^2 {\bf v}$, should, in general, be complex \cite{irgens}. However, this coefficient varies with the excitation frequency, with its real part, which governs conventional viscous dissipation, becoming negligible at low frequencies. Therefore, the term $i \gamma \nabla^2 {\bf v}$ can be seen as a low-frequency approximation of a full viscoelastic response, becoming increasingly accurate for long wavelength phonons.

The momentum storage acts a driving force to return the fluid micro-structure to its equilibrium state. This is a very specific kind of driving originated from viscoelastic effect. A conceptually simpler form of elastic restoring force is that typical of a spring, governed by the equation of motion $\dot{\psi}=i \gamma^{'} \psi$, where $\gamma^{'}$ is the natural frequency of the oscillation. 

By replacing the storage viscosity term $i \gamma \nabla^2 \psi_1$ with the restoring force term $\Phi_1=i \gamma^{'} \psi_1$ in Eq. (\ref{eq11}) the form of the phonon dispersion relation changes from DSR1 to DSR2. Indeed, using the previous expressions for $P_1$ and $\phi_1$, we get
\begin{equation}
\partial_{TT}^2 \psi_1 = c_s^2 \nabla^2 \psi_1   - \Omega_0^2 \psi_1  - i \gamma^{'} \partial_T \psi_1 \, .
\label{eq13}
\end{equation}
which leads to the dispersion relation 
\begin{equation}
E^2 = c_s^2 p^2 + m^2 c^4 \left(1 - \frac{\gamma^{'} E}{m^2 c_s^4}\right)
\label{eq14}
\end{equation}
Defining $E_p= 2 m^2 c_s^4/\gamma^{'}$ and rescaling the sound speed and phonon mass as
\begin{equation}
c_s^2 \rightarrow \frac{c_s^2}{\left(1 - \frac{m^2 c_s^4}{E_p} \right)} 
\,\,\,\,\,\,\, m^2 \rightarrow \frac{m^2}{\left (1 - \frac{m^2 c_s^4}{E_p}\right)} \,\,\,
\label{eq15}
\end{equation}
Eq. (\ref{eq14}) reduces to the DSR2 energy-momentum relation (\ref{eq2}). The rescaling (\ref{eq15}) are equivalent to assume a new barotropic law $P^{'}=(1 + \gamma^{'}/4 \beta \rho_0) P$ and to define the rest frequency as $\Omega_0^2 = \beta \rho_0 + \gamma^{'}/4$. Accordingly, the Planck frequency can be expressed as $\Omega_p = (4 \beta \rho_0 + \gamma^{'})/2 \gamma^{'}$.

\section{Discussion and conclusions}

Acoustic analogue gravity models have been shown to provide a useful testbed for field theory in curved spacetime. Beyond offering experimental verification of fundamental phenomena that might otherwise remain unobserved, these systems have also contributed to a deeper theoretical understanding of such effects. 

We have shown that by tailoring the interaction potentials, we can explicitly construct hydrodynamic models in which phonons propagate accordingly to doubly special relativistic kinematics, with the sound speed playing the role of the speed of light and the analogue of Planck scale arising from viscoelastic processes. In particular we derived models for acoustic analogue simulations of DSR1 and DSR2 theories.

For clarity, we report below the full hydrodynamic equations of the two models
\begin{eqnarray}
\partial_t \rho + \nabla\cdot(\rho {\bf v}) = 0 \, \\
\partial_t {\bf v} + ({\bf v} \cdot {\bf \nabla}) {\bf v}= - \frac{\nabla P}{\rho} - \nabla \phi - \nabla \Phi  \, \\
\nabla^{2} \phi = -\beta \rho
\label{eq16}
\end{eqnarray}
where $P=P(\rho)$ is a generic barotropic law, $\phi$ is a repulsive dipolar potential and $\Phi$ accounts for viscoelatic driving forces. 

When $\Phi=i \gamma \nabla^2 \psi$ (elastic storage), phonons displays the $\mathcal{O}(E^3/E_p^3)$ approximation of Amelino-Camelia DSR1 dispersion relation. When $\Phi=i \gamma^{'} \psi$ (elastic restoring), they follow the Magueijo-Smolin model.

The models here discussed present a certain degree of idealization, in particular for the absence of losses and an oversimplified description of viscoelastic effects. However, our goal was not to outline an experimental proposal, but rather to demonstrate that DSR scenarios can naturally arise in the kinematics of collective excitations within physically consistent fluid models. This suggest the possibility that other analogous theories could be simulated using different interaction potentials, enabling comparisons of the hydrodynamic mechanisms that originate them. On the other hand, we expect that these results might also trigger the interest of experimentalists both in classical and quantum fluids where interactions have recently been engineered and tuned with unprecedented precision, offering exciting prospects for analogue gravity simulations.

\subsection{Acknowledgements}

I acknowledge M. Ciszak for carefully reading the manuscript

\end{document}